\documentclass[preprint]{aastex}

\def\araa{ARA\&A}             % Annual Review of Astron and Astrophys
\def\apj{ApJ}                 % Astrophysical Journal
\def\apjl{ApJ}                % Astrophysical Journal, Letters
\def\apjs{ApJS}               % Astrophysical Journal, Supplement
\def\aap{A\&A}                % Astronomy and Astrophysics
\def\mnras{MNRAS}             % Monthly Notices of the RAS
\usepackage{graphicx}

\shorttitle{Transient LMXBs in NGC\,3379 and NGC\,4278}
\shortauthors{T. Fragos et al.}

\begin{document}

\title{TRANSIENT LOW-MASS X-RAY BINARY POPULATIONS IN ELLIPTICAL GALAXIES NGC\,3379 AND NGC\,4278} 

\author{T.\ Fragos$^{1}$, V.\ Kalogera$^{1}$, B.\ Willems$^{1}$, K.\ Belczynski$^{2}$, G.\ Fabbiano$^{3}$, N.\ J.\ Brassington$^{3}$, D.-W.\ Kim$^{3}$,  L.\ Angelini$^{4}$, R.\ L.\ Davies$^{5}$, J.\ S.\ Gallagher$^{6}$, A.\ R.\ King$^{7}$, S.\ Pellegrini$^{8}$, G.\ Trinchieri$^{9}$, 
S.\ E.\ Zepf$^{10}$, A.\ Zezas$^{3}$} 

\altaffiltext{1}{Northwestern University, Department of Physics and Astronomy, 2145 Sheridan Road, Evanston, IL 60208, USA}
\altaffiltext{2}{New Mexico State University, Department of Astronomy, 1320 Frenger Mall, Las Cruces, NM 88003, USA}
\altaffiltext{3}{Harvard-Smithsonian Center for Astrophysics, 60 Garden Street, Cambridge, MA 02138}
\altaffiltext{4}{Laboratory for High Energy Astrophysics, NASA Goddard Space Flight Center, Code 660, Greenbelt, MD 20771}
\altaffiltext{5}{Denys Wilkinson Building, University of Oxford, Keble Road, Oxford OX1 3RH, UK}
\altaffiltext{6}{Astronomy Department, University of Wisconsin, 475 North Charter Street, Madison, WI 53706}
\altaffiltext{7}{University of Leicester, Leicester LE1 7RH, UK}
\altaffiltext{8}{Dipartimento di Astronomia, Universita` di Bologna, Via Ranzani 1, 40127 Bologna, Italy}
\altaffiltext{9}{INAF Observatorio Astronomico di Brera, Via Brera 28, 20121 Milan, Italy}
\altaffiltext{10}{Department of Physics and Astronomy, Michigan State University, East Lansing, MI 48824-2320}

\email{tassosfragos@northwestern.edu, vicky@northwestern.edu, kbelczyn@nmsu.edu,  gfabbiano@cfa.harvard.edu, nbrassington@head.cfa.harvard.edu, kim@cfa.harvard.edu,  angelini@davide.gsfc.nasa.gov, rld@astro.ox.ac.uk,  jsg@astro.wisc.edu, ark@star.le.ac.uk, silvia.pellegrini@unibo.it, ginevra.trinchieri@brera.inaf.it, zepf@pa.msu.edu,  azezas@cfa.harvard.edu}

\begin{abstract}

We propose a physically motivated and self-consistent prescription for the modeling of transient neutron star (NS) low-mass X-ray binary (LMXB) properties, such as duty cycle (DC), outburst duration and recurrence time. We apply this prescription to the population synthesis (PS) models of field LMXBs presented by \citet{Fragos2008}, and compare the transient LMXB population to the \emph{Chandra} X-ray survey of the two elliptical galaxies NGC\,3379 and NGC\,4278, which revealed several transient sources \citep{Brassington2008,Brassington2009}. We are able to exclude models with a constant DC for all transient systems, while models with a variable DC based on the properties of each system are consistent with the observed transient populations. We predict that the majority of the observed transient sources in these two galaxies are LMXBs with red giant donors. Our comparison suggests that LMXBs formed through evolution of primordial field binaries are dominant in globular cluster (GC) poor elliptical galaxies, while they still have a significant contribution in GC rich ones.

\end{abstract}


\maketitle

\section{INTRODUCTION}

Recent \emph{Chandra} observations \citep{Kim2006,Brassington2008,Brassington2009} have yielded the first low-luminosity X-ray luminosity functions (XLF) of LMXBs for two typical old elliptical galaxies, NGC\,3379 and NGC\,4278. The detection limit in these observations was $\sim 3\times10^{36}\rm \,erg\,s^{-1}$, an order of magnitude lower than in most previous similar surveys, while the completeness limit was  $\sim 10^{37} \rm \,erg\,s^{-1}$. In a followup study, \citet{Brassington2008,Brassington2009} used  multiple \emph{Chandra} pointings to identify potential transient sources. In that work, a \emph{transient candidate} was defined as a source with X-ray luminosities varying by at least a factor of 10, above the completeness limit, between different pointings. Similarly a \emph{potential transient candidate} was defined as a source with X-ray luminosities varying by at least a factor of 5. Using these definitions, 5 transient candidates and 3 additional potential transient candidates were identified in NGC\,3379, while for NGC\,4278 the corresponding numbers were 3 and 3 respectively.

\citet{Piro2002} conjectured that LMXBs in the field of ellipticals must be dominated by NS accreting from red giants. Using analytical approximating formulae for the evolution of binary systems they estimated that such binaries are expected to be transient, with maximum recurrence times of $100 - 10000\rm\, yr$, for at least 75\% of their lifetimes. However, they limited their study only to LMXBs with red giant donors, and were unable to calculate the duration of the outburst phase or to predict how many transients would be identified in repeated \emph{Chandra} visits.

\citet{Fragos2008} presented results from extensive LMXB population synthesis (PS) simulations for NGC\,3379 and NGC\,4278. They considered models for the formation and evolution of LMXBs from primordial and isolated field binaries with different common envelope efficiencies, stellar wind prescriptions, magnetic braking laws, and initial mass functions. They identified models that produce XLFs consistent with observations both in shape and normalization, suggesting that a primordial galactic field LMXB population can have a significant contribution to the total population of an elliptical galaxy. 

In this letter we expand on the work by \citet{Fragos2008}, and propose a physically motivated and self-consistent prescription for the modeling of transient LMXB properties. Furthermore, we carry out Monte Carlo simulations to compare our findings to the recent observational work by  \citet{Brassington2008,Brassington2009}. In \S~2 we describe briefly the PS models we use for the formation and evolution of  field LMXBs . In \S~3 we expand on the detailed modeling of  transient LMXBs and our comparison of PS models with observations. We show our results  in \S~4, and discuss the implication of our findings for transient LMXBs in \S~5.

\section{LMXB POPULATION SYNTHESIS MODELS}

The models presented in the study by \citet{Fragos2008} were focused on  LMXBs formed in the galactic field as products of the evolution of isolated primordial binaries. The simulations were performed using \textit{StarTrack} \citep{BKB2002, Belczynski2008}, an advanced PS code that has been tested against and calibrated using detailed MT calculations with stellar structure and evolution codes, and observations of binary populations. In the development of these models, all current knowledge about the stellar population in these galaxies were incorporated \citep[see Tables 1 and 2 in ][]{Fragos2008}. The observationally determined parameters, such as their age ($9.5-10.5\,\rm Gyr$) and metallicity ($[Fe/H]=0.14-0.16$), or their total stellar mass ($8.6\times10^{10}-9.4\times10^{10}\,\rm M_{\odot}$), are similar for NGC3379 and NGC4278. However, there are physical processes involved in the formation and evolution of a binary system, such as stellar winds, initial mass function, common envelope efficiency, magnetic braking and transient behavior, which are not fully understood quantitatively. Therefore various prescriptions were used to model them.  A total of 336 models were studied. See Tables 3 and 4 in \citet{Fragos2008} for all the model parameters and naming conventions.

\citet{Fragos2008} found that a small subset of their models produce XLFs in very good agreement with the observations, based on both the XLF shape and absolute normalization. They concluded that formation of LMXBs in the galactic field via evolution of primordial binaries \emph{can} have a significant contribution to the total population of an elliptical galaxy, especially the GC poor ones such as NGC3379 \citep{Fabbiano2007}. In our analysis here, we adopt their best-fitting model (hereafter model $14^{IT}$), which produces the XLF in best agreement with observations. Model $14^{IT}$ assumes a moderate common envelope efficiency $\alpha_{CE}=0.5$, a standard stellar wind prescription prescription \citep[as described in ][]{HTP2002}, the magnetic braking law derived by \citet{IT2003}, and a Salpeter initial mass function. 

The LMXB population predicted by model $14^{IT}$ has a significant contribution from transient systems, which with reasonable outburst DC can even dominate the XLF. As a consequence the XLF shape is rather sensitive to the treatment of these transient systems. \citet{Fragos2008} showed that keeping the same PS parameters and changing \emph{only} the modeling of transient sources leads to different XLFs. They tried different methods of modeling the outburst characteristics of transient LMXBs and got the best agreement with observations when they consider a variable DC for NS LMXB based on the binary properties, hereafter model $14A^{IT}$. However, a constant low DC ($\sim 1 \%$), hereafter model $14B^{IT}$, was also statistically consistent with the observed XLF. In both cases, the outburst luminosity was calculated as described in section 3.1.

\section{MODELING THE TRANSIENT BEHAVIOR OF LMXBS}
\subsection{Duty Cycle of Transient X-ray Binaries}

In our models we keep track of all binary properties, including the MT rates, as a function of time for all accreting compact objects. Here we focus only on NS LMXBs, which dominate the total population \citep[see Fig.~2 in ][]{Fragos2008}. We use the MT rate $\dot{M}_{\rm d}$ to identify transient sources in our simulation results. In the context of the thermal disk instability model, mass transferring binaries with MT rate lower than the critical rate $\dot{M}_{\rm crit}$ for the thermal disk instability to develop \citep{Paradijs1996,KKB1996,DLHC1999,MPH2002},  are considered transient sources, meaning that they spend most of their life in quiescence ($T_{\rm quiescence}$), when they are too faint to be detectable, and they occasionally go into an outburst. The fraction of the time that they are in outburst ($T_{\rm outburst}$) defines their DC: 
\begin{equation}{\rm DC} \equiv T_{\rm outburst}/\left(T_{\rm outburst}+T_{\rm quiescence}\right). \label{dc_def}\end{equation}

The details of the thermal disk instability model are not well understood therefore the values of the outburst luminosity and the DC  cannot be calculated directly from first principles. A simple but physically motivated treatment is to assume that in the quiescent state the NS does not accrete any (or accretes an insignificant amount of) mass, and matter from the donor is accumulated in the disk. In the outburst state all this matter is accreted onto the NS emptying the disk. Taking into account also that the X-ray luminosity probably cannot exceed $L_{\rm Edd}$ by more than a factor of 2 \citep[cf.][]{TCS1997}, we define the outburst luminosity as:
 \begin{equation} L_{\rm x} = \eta_{\rm bol}\epsilon\times
 \min{\left(2\times L_{\rm Edd} , \frac{GM_{\rm a}\dot{M}_{\rm d}}{R_{\rm a}} \times \frac{1}{\rm DC}\right)}, \label{Lx_phys} \end{equation}
where $M_{\rm a}$ and $R_{\rm a}$ are the mass and the radius of the accretor, $dot{M}_{\rm d}$ is the mass-loss rate of the donor and $\eta_{\rm bol}$ is correction factor that converts the bolometric luminosity to the observed \emph{Chandra} band.

In equation~\ref{Lx_phys}, DC is unknown. \cite{DLM2006} studied accretion disk models for dwarf novae which are thought to experience the same thermal disk instability. They found a correlation between the DC of the system and the rate at which the donor star is losing mass $\dot{M}_{\rm d}$. The exact relation between these quantities depends on the disk's viscosity parameters, but the general behavior can be approximated by:
\begin{equation} DC\approx \left(\dot{M}_{\rm d}/\dot{M}_{\rm crit} \right)^{2}. \label{dc} \end{equation}
Plugging equation~(\ref{dc}) into equation~(\ref{Lx_phys}) we eliminate the DC dependence and get an expression that depends only on quantities which are directly calculated in our PS modeling.

\subsection{Recurrence Time and Duration of X-ray Outbursts}

The derivation of the critical MT rate $\dot{M}_{\rm crit}$ for the thermal disk instability, comes from requiring that some part of the accretion disc has a temperature below the hydrogen ionization temperature. In the thin disk approximation the  effective temperature decreases with increasing disk radius, so it is sufficient to require that the effective temperature at the outer edge of the disk is below the hydrogen ionization temperature. In X-ray binaries with NS accretors, the surface temperature of the disk is regulated by irradiation from the accretor \citep{Paradijs1996,DHL2001}.

Following the analysis presented by \citet{Piro2002}, we assume that the accretion disk around the NS extends up to a radius of 70\% of the Roche lobe radius of the NS ($R_{\rm RL,1}$), and that a transient NS LMXB is in the quiescent state until the surface density at the edge of the disk reaches the critical surface density 
 
\begin{equation} \Sigma_{\rm max} = 644\left(M_1/M_{\odot}\right)^{-0.37}\left(R/R_\odot\right)^{1.11}\rm{\,g\,cm^{-2}} \label{Smax} \end{equation} 
that will lead to instability \citep{DHL2001}. In equation~\ref{Smax} we assume a pure hydrogen disk with  viscosity parameter $\alpha = 0.1$, and no irradiation of the disk while the binary is in the quiescent state. For systems with helium accretion disks, such as LMXB with white dwarf donors, we modify equation~\ref{Smax} accordingly \citep[see equation~2 in][]{LDK2008}. Hence, the maximum mass accumulated in the accretion disk, before the X-ray binary goes into outburst, is:
\begin{equation}  M_{\rm disk, max} = \int^{0.7 R_{\rm RL,1}}_{R_{\rm NS}} \Sigma_{\rm max}(R)2\pi R dR \end{equation} 
From the equation above we can set an upper limit on the duration of the quiescent phase and the outburst. Using the mass-loss rate of the secondary star, as given by our PS models, and equations \ref{dc_def} and \ref{dc}, the time duration of the quiescent and outburst phase becomes:
\begin{equation} T_{\rm quiescence} = \frac{M_{\rm disk, max}}{\dot{M}_{\rm d}}, \ \ T_{\rm outburst} = \frac{\dot{M}_{\rm d}^2}{\dot{M}_{\rm crit}^2-\dot{M}_{\rm d}^2} M_{\rm disk, max}.\end{equation}

\subsection{Detection probability of transient LMXBs}

\begin{figure}
\plotone{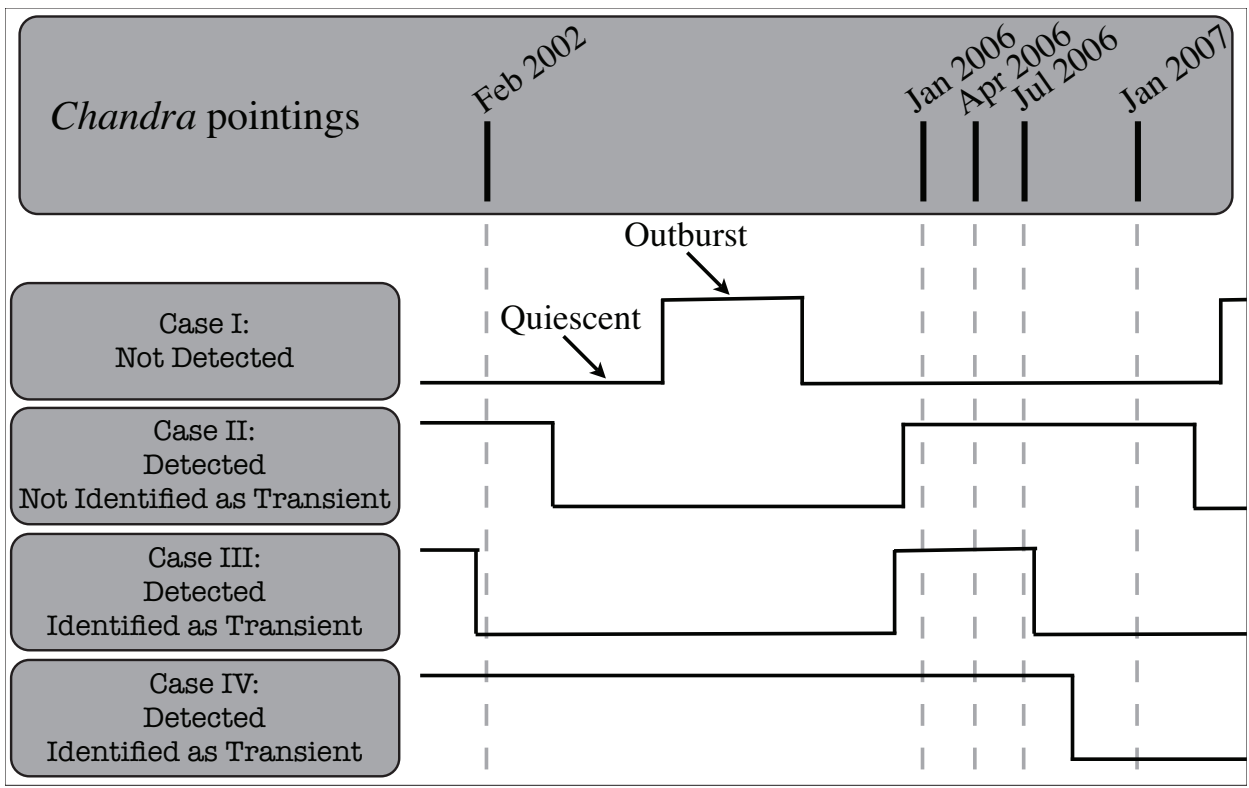}
\caption{Schematic description of the procedure followed to calculate the probability that a LMXB is identified as a transient source in a survey.}
\label{trans_fig}
\end{figure}

The probability that a transient LMXB is detected as a bright X-ray source is equal to its DC. However, the probability that a LMXB is identified as a transient source in a survey with multiple observations depends on the duration and the recurrence time of the outbursts of the source, as well as the time spacing between the observations. In our analysis we assume that a transient LMXB can be in two states, outburst or quiescent, and that its luminosity in outburst is given by equation~\ref{Lx_phys}, while in quiescence it is undetectable (i.e. we do not take into account rise and decay characteristics of the outburst). Hence, in order for a source to be identified as a transient, it has to be found at a different state, in at least one out of the multiple observations (see Figure~\ref{trans_fig}). 

Given a model population of LMXBs, we calculate for each transient LMXB the probability that it would be identified as a transient source in an observational survey like the ones for the elliptical galaxies NGC\,3379 and NGC\,4278 \citep{Kim2006}. In order to estimate this probability, we first calculate the  outburst and recurrence time for the specific binary as described in section 3.2, and then create an artificial lightcurve. This lightcurve should include at least one full cycle of the transient and be at least twice as long as the time interval between the first and the last observation. Next, we choose randomly a time in this artificial light curve to position the first pointing of the survey with the other pointings following according to the time intervals between the pointing of the survey we are simulating. For this random positioning of the first pointing, we attain the state of the transient LMXB (outburst or quiescent) at each of the pointings, and assess whether this system would be identified as a transient source (see Figure~\ref{trans_fig}). We repeat the procedure $10^5$ times, choosing randomly a different starting time in the artificial lightcurve, which enable us to estimate the probability that a specific system would be identified as a transient source.   
	
Based on this probability for each transient LMXB of the model population, we perform another Monte Carlo simulation, this time to estimate the total number of transients that would be identified in the modeled survey. Repeating this procedure $10^5$ times, we are able to derive a probability density function (PDF) of the number of identified transient sources, for a combination of PS and X-ray survey parameters.

\section{RESULTS}

As mentioned earlier,  we adopt the PS model $14^{IT}$, which was chosen based on comparison of the shape of the modeled XLF to the observed one \citep{Fragos2008}. We examine two prescriptions for the transient characteristics of LMXBs: a variable DC based on the properties of each binary as described in section 3.1 (model $14A^{IT}$), and a low constant DC of 1\% (model $14A^{IT}$). For more details on the model parameters see sections 2.3 and 3.1 in \citet{Fragos2008}. We normalize the total number of LMXBs in our models in two different ways: based on the total stellar mass of the galaxy, or based on the number of observed LMXBs above the completeness limit of the observations. 

In order to simulate the observational identification of transient sources as realistically as possible, for each of the two elliptical galaxies we use the actual time intervals between the successive pointings of the survey. Furthermore, we try to mimic  the definitions of transient candidate and potential transient candidate of \citet{Brassington2008,Brassington2009}. In our simulated observations, a  transient candidate is a system that is identified as transient and has outburst luminosity above $10^{38}\rm \, erg\, s^{-1}$, while a potential transient candidate has outburst luminosity between $5\times 10^{37}\rm \, erg\, s^{-1}$ and $10^{38}\rm \, erg\, s^{-1}$.

\begin{figure}
\plotone{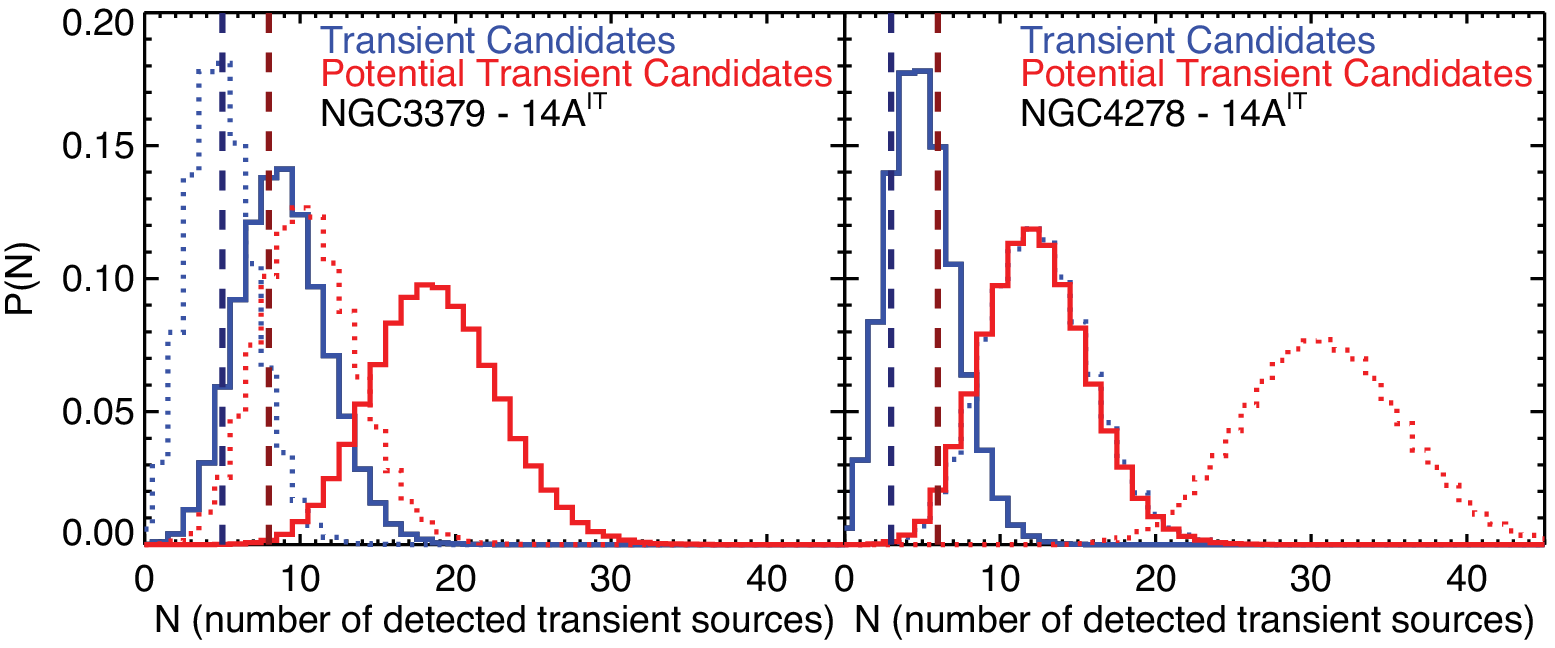}
\plotone{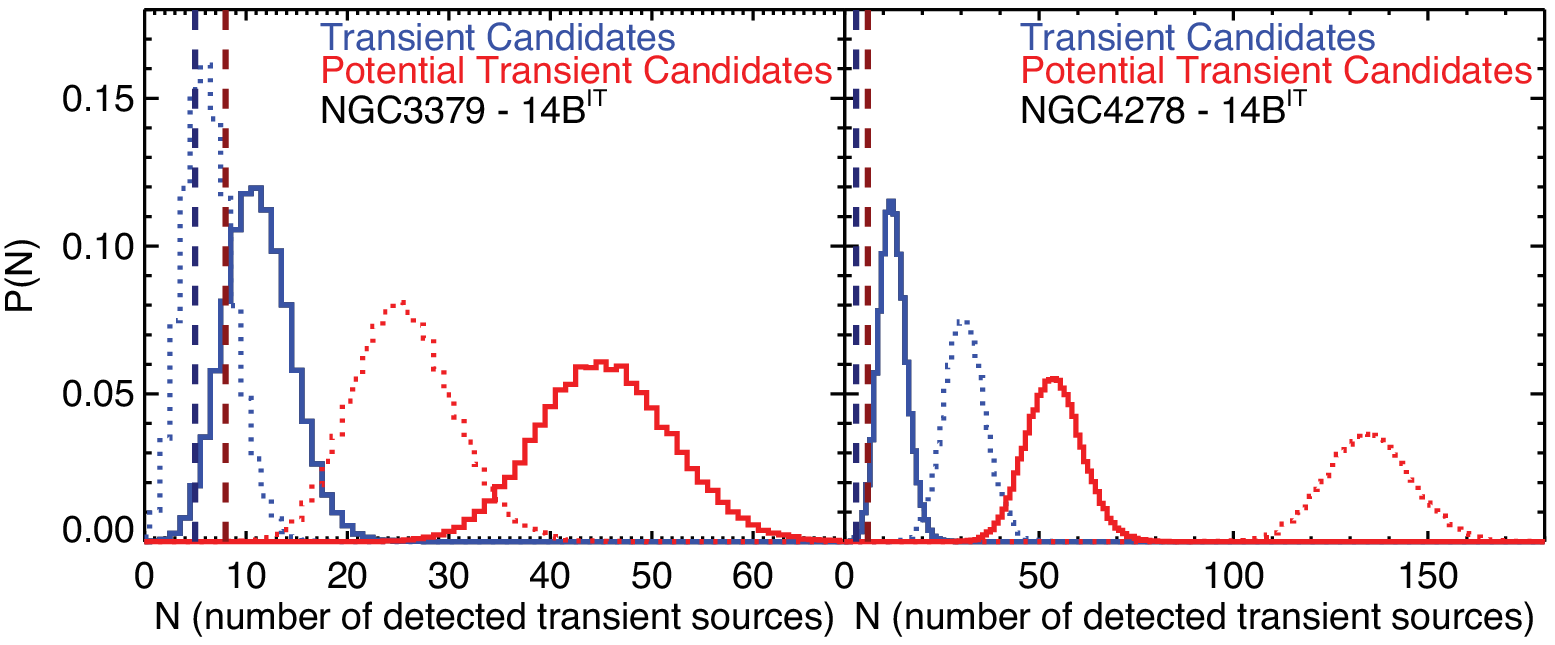}

\caption{ PDF of the number of detected transient sources in an X-ray survey such as that by \citet{Brassington2008,Brassington2009}, for model $14A^{IT}$ in panels (a) and (b) and model $14B^{IT}$ in panels (c) and (d). Panels (a) and (c) correspond to the elliptical galaxy NGC\,3379, while panels (b) and (d) refer to NGC\,4278. Solid lines are used when the normalization of the total number of LMXBs in our models is based on the total stellar mass of the galaxy, whereas dotted lines signify a normalization based on the observed number of LMXBs above the detection limit. The vertical dashed lines refer to the actual observations by \citet{Brassington2008,Brassington2009}}
\label{transients}
\end{figure}

Figure~\ref{transients} shows the  PDF for the number of detected sources in an X-ray survey such as that by \citet{Brassington2008,Brassington2009}. When we normalize the total number of LMXBs to the stellar mass of the the galaxy, model $14A^{IT}$ (variable DC) gives  PDFs consistent with observations for both transient candidates and potential transient candidates, and for both galaxies. On the other hand, model $14B^{IT}$ (constant DC $\sim 1\%$) is consistent with observations only when we look at the bright transient sources (i.e., transient candidates). A constant DC significantly  overproduces the number of potential transient candidates.

Normalizing the total number of LMXBs of our models to the observed number of LMXBs above the completeness limit of the observation ($\sim  10^{37}\rm \, erg\, s^{-1}$), for each of the two galaxies separately, alters our findings. In the case of NGC\,3379, and for model $14A^{IT}$, the  PDFs of both transient candidates and potential transient candidates are now in excellent agreement with the observations. This suggest that our models slightly overproduce LMXBs. This was already mentioned in \citet{Fragos2008}, where the authors explained that  the PS model normalization can be very sensitive to certain PS parameters, such as the distribution of the mass ratios between the initial masses of the two binary components. Without a proper multi-dimensional coverage of the parameter space of at least thousands of models and given the uncertainties in the total galaxy mass, we cannot use the total number of LMXBs as a formal constraint to a factor better than a few. For the case of NGC\,4278 on the other hand, normalizing our model population to the number of observed LMXBs results to a very high number of expected identified transient source, which is inconsistent with observations. However, this does not come as a surprise, since the elliptical galaxy NGC\,4278 has a large number of GC ($\sim 5$ times more than NGC\,3379) and a significant fraction of the observed LMXB population resides in GC. It has already been suggested that luminous GC LMXBs are  predominantly persistent source \citep{BD2004, Ivanova2008} in which case the similar number of identified transients in the two elliptical galaxies, despite the difference by a factor of $\sim 2$ in the total number of observed LMXBs, is expected.

\begin{figure}
\plotone{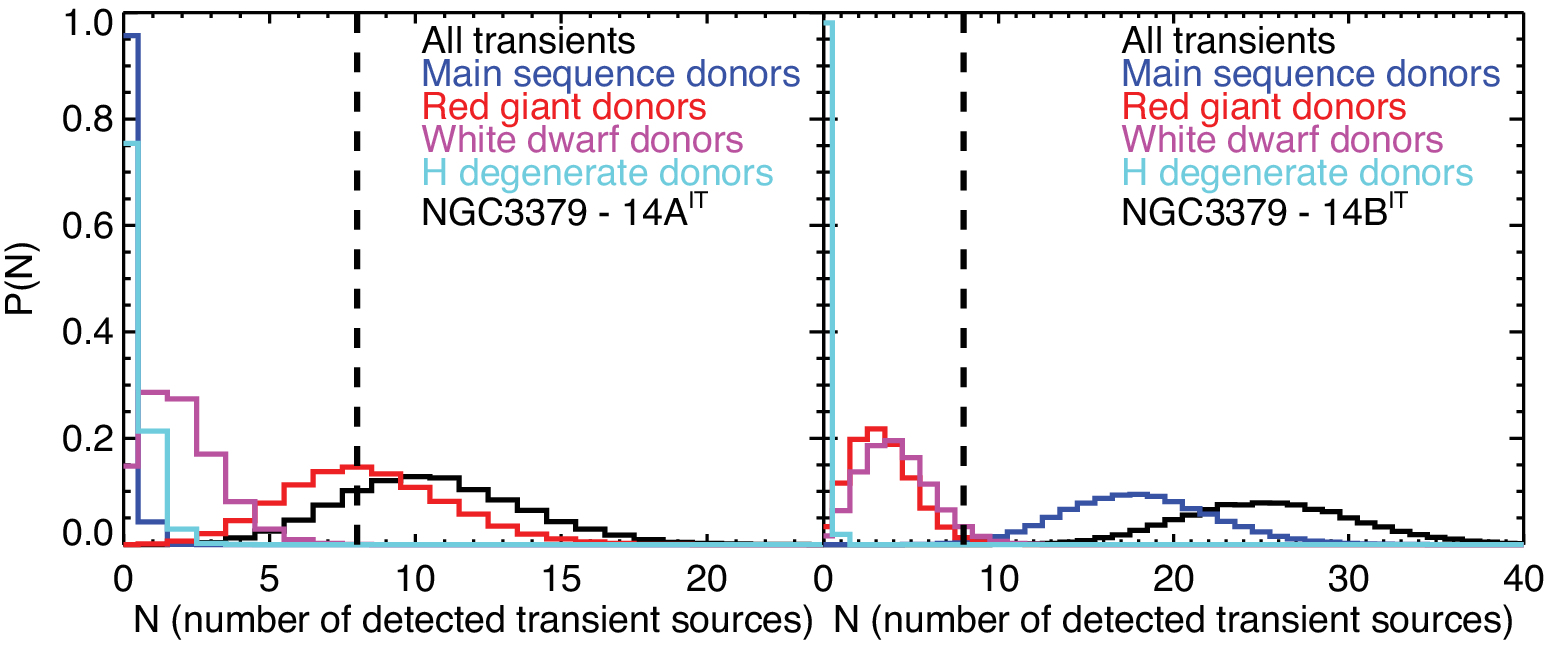}
\caption{ PDF of the number of potential transient candidates for the model $14A^{IT}$ (variable DC) and model $14B^{IT}$ (constant DC). We show the contribution of different sub-populations by separating the LMXBs into groups of systems with different donor. The vertical dashed lines refer to the observed number.}
\label{transients_split}
\end{figure}

\citet{Fragos2008} analyzed their model LMXB populations and found that the sub-populations that mainly contribute to the model XLFs are transient and persistent NS-LMXBs with red giant donors. Following a similar analysis, in Figure~\ref{transients_split} we show again the  PDF of the number of potential transient candidates for models $14A^{IT}$ and $14B^{IT}$ of the elliptical galaxy NGC\,3379, but this time we split the modeled population into sub-populations based on the type of the donor star. We find that in model $14A^{IT}$, the majority of identified transient sources are LMXBs with red giant donors, as already argued in the literature \citep{KKB1996,Piro2002, Fragos2008}. In contrast, model $14B^{IT}$ predicts a much higher than the observed number of identified transient LMXBs and the majority of them have main sequence donors. This results from assigning the same low DC to all systems, whereas transient LMXBs with main sequence donors have usually MT rates close to $M_{\rm crit}$. These systems would otherwise have $DC\approx 1$ and low outburst luminosity, and thus would be undetectable at X-ray luminosities above $5\times 10^{37}\rm \, erg\, s^{-1}$.

\section{DISCUSSION}

The goal of this work was to pose further constraints on our models based on the recent observational work by \citet{Brassington2008,Brassington2009}, and simultaneously gain a better understanding on the nature of transient LMXBs. We found that the populations of transient LMXBs produced by the PS model $14A^{IT}$ \citep{Fragos2008} is consistent with the observed population. In this model a variable DC was assigned for each system based on its properties (see section 3.2). In contrast, we show that the widespread in the literature assumption of a constant low ($\sim 1\%$) DC for all transients, although it can result in a XLF consistent with the observed one \citep{Fragos2008}, is not consistent with the number of observationally identified transient LMXBs.

While the two elliptical galaxies modeled here have very similar properties in terms of stellar mass, age, star formation history and metallicity, the major difference between the two is their specific GC frequency. NGC\,4278 has about 5 times more GC per unit mass compared to NGC\,3379. \citet{Kim2006} using similar \emph{Chandra} observing times for both galaxies, found that NGC\,4278 has $\sim 3$ times more LMXBs with luminosity above $10^{37}\rm \, erg\, s^{-1}$ than NGC\,3379. The excess of X-ray sources in NGC\,4278 can be attributed to the higher LMXB formation efficiency in GC, suggesting that for the GC-rich elliptical galaxies the formation of LMXBs in GC through dynamical interactions plays an important role. At the same time, the number of identified transient LMXBs is roughly the same in both elliptical galaxies. Despite the small statistical sample of only two galaxies, it seems that the number of transient sources is proportional to the stellar mass of the galaxy rather than the number of its GC. 

The same argument is also supported by our PS models which only take into account LMXB formation through binary evolution of field primordial binaries. When we normalize the modeled LMXB population to either the stellar mass of the galaxy or to the number of observed LMXBs in the GC-poor galaxy NGC\,3379, the predicted number of observable transient LMXBs is consistent with observations \citep{Brassington2008,Brassington2009}. Hence, we conclude that LMXBs formed through the evolution of primordial field binaries dominate the observed LMXB populations in GC-poor elliptical galaxies, while in GC-rich ellipticals the formation of LMXB through dynamical interactions in GC becomes more important. However, even in the latter case, the primordial field population has a significant contribution.   

\acknowledgements 
This work was supported by Chandra G0 grant G06-7079A and subcontract G06-7079B.

\end{document}